# Shear Mediated Elongational Flow and Yielding in Soft Glassy Materials

Asima Shaukat, Manish Kaushal, Ashutosh Sharma and Yogesh M. Joshi[*]

Department of Chemical Engineering,

Indian Institute of Technology Kanpur, Kanpur 208016 INDIA

* Corresponding Author, E- mail: joshi@iitk.ac.in,

Tel.: +91-512-2597993, Fax: +91-512-2590104

**Abstract**

In this work, we study the deformation behavior of thin films of various soft glassy materials that are simultaneously subjected to two creep flow fields, rotational shear flow by applying torque and elongational flow by applying normal force. The generic behavior under the combined fields is investigated in different soft glassy materials with diverse microstructure such as: hair gel, emulsion paint, shaving foam and clay suspension. Increase in strength of one stress component while keeping the other constant, not only leads to an expected enhanced deformation in its own direction, but also greater strain in the other direction. Herschel-Bulkley model is observed to explain this behavior qualitatively. Elongational flow induced in the materials eventually causes failure in the same. Interestingly time to failure is observed to be strongly dependent not just on normal force but also on the applied rotational shear stress. We believe that the presence of a three dimensional jammed structure, in which overall unjamming can be induced by applying stress having sufficient magnitude irrespective of the direction leads to the observed behavior. In addition, we observe self-similarity in the elongational as well as rotational strain – time curves corresponding to various combinations of both the fields. This observation suggests a mere shift in the time-scales involved keeping the path followed in the process unchanged. A phase diagram is also constructed for various soft glassy materials by determining different combinations of orthogonal stresses beyond which materials yield. Estimated yield stress in the limit of flow dominated by applied tensile force on the top plate demonstrates scatter, which might be originating from fingering instability. Except this deviation, yielding is observed when the invariant of stress tensor exceeds yield stress, validating the Von Mises criterion.

**Keywords:** Soft glassy materials, yielding, Von Mises criterion, elongational flow.

## I.  Introduction:

In soft materials such as concentrated colloidal suspensions and emulsions, colloidal gels, foams, etc., the constituent entities/particles are trapped within the cages formed by the surrounding particles leading to a jammed state. Such systems do not explore the total available phase space in the experimental time scales because of the constraints on the translational mobility of the constituents.[1, 2] In order to lower the energy, these out–of–equilibrium systems undergo structural evolution with respect to time.[1-12] The microscopic dynamics of structural rearrangement in these materials gets strongly affected by temperature [13-16] and stress/strain fields.[5, 12, 17-27] Application of a stress field on such materials leads to partial or complete un-jamming depending on the yield stress of the material.[20] Typically a stress field imparts energy to the particles which facilitates their diffusion out of their cages. Under small stresses an elastic response is observed, on the other hand for stress levels greater than the yield stress, a structural breakdown results in a plastic flow.[5, 17, 20, 22, 28-35] However, owing to the presence of shear localization and thixotropy of the material, an accurate determination of the yield stress is a difficult exercise.[36-38] In addition, the yield stress of soft glassy materials, like other rheological properties, depends upon the deformation history as well as the time elapsed since mechanical quench.[10, 20, 39-41] A general constitutive relation for a yield stress fluid can be written as follows:[42, 43]

$$\underset{\sim}{\tau} = E\underset{\sim}{\gamma}, \qquad \text{for } \sqrt{\underset{\sim}{\tau}:\underset{\sim}{\tau}/2} < \tau_y$$

$$\underset{\sim}{\tau} = \left(\frac{\tau_y}{\dot{\gamma}} + \mu\right)\underset{\sim}{\dot{\gamma}}, \quad \text{for } \sqrt{\underset{\sim}{\tau}:\underset{\sim}{\tau}/2} \geq \tau_y \qquad (1)$$

where $\underset{\sim}{\tau}$ is deviatoric stress tensor, $\underset{\sim}{\gamma}$ is strain tensor, $E$ is elastic modulus, $\underset{\sim}{\dot{\gamma}}$ is rate of strain tensor, $\tau_y$ is the yield stress, $\dot{\gamma}$ is the second invariant of rate of strain tensor $\left(\sqrt{\underset{\sim}{\dot{\gamma}}:\underset{\sim}{\dot{\gamma}}/2}\right)$, and $\mu$ is some function of $\dot{\gamma}$. For a constant $\mu$ equation (1) leads to Bingham model; while for a power law dependence of $\mu$ on $\dot{\gamma}$, we get Herschel-Bulkley model: $\mu = m\dot{\gamma}^{n-1}$, where $m$ and $n$ are model parameters.[44] The criterion that yield stress needs to be greater than the invariant of stress tensor for flow to take place is also called Von Mises criterion.[45, 46] The ability to undergo flow when deformed and to behave as solids otherwise makes these materials very desirable for use in cosmetic, pharmaceutical, chemical and food industries. Some examples are shaving foams, hair gels, tooth pastes, creams, lotions, jams, jellies, paints, etc. However, their high viscosities and plastic flow behavior poses challenges in their processing.

Most of the previous studies related to these materials are focused on their behavior under application of simple one directional stress or strain fields. These include studies under pure rotational shear flow fields,[17, 20] elongational flow[23, 47-51] and squeeze flow fields.[52-55] Generally, the principle focus of such studies is to determine the response in terms of the strain/compliance induced in the material as a function of time under mentioned loading conditions. However, there are numerous applications where more than one stress field act simultaneously on soft glassy materials. For example, in the case of lubricant films, hip joints, spinning of fibers, drilling of mud, polymer extrusion, droplet break-up, spraying, atomization, pressure sensitive adhesives, spiral vortex flow under axial sliding, etc.[53, 56-58]

Recently Brader and coworkers[45] proposed a single mode MCT (mode coupling theory) model having a tensorial structure and presented the dynamic yield surface as a function of various combinations of components of stress tensors. Their estimated yield surface, which separates the solid and liquid region, closely matched Von Mises criterion. On the other hand, on an experimental front Ovarlez and co-workers[46] performed squeeze flow experiments combined with rotational shear flow for different soft jammed materials and observed that if a material gets un-jammed in one direction under a flow field, it simultaneously yields in all the other directions. Moreover, they verified Von Mises criterion experimentally for the first time for soft jammed materials. They also observed that the overall viscous response of the material is always dominated by the primary flow induced in the material. Interestingly MCT approach of Brader and coworkers captured this behavior very well.[59]

In this work, we study the deformation behavior of variety of soft glassy materials under simultaneous application of two stress fields orthogonal to each other, namely a normal (tensile) force field and a rotational shear flow field. A tensile flow field, though similar to squeeze flow field with exception of the direction of the applied force, has some significant differences from the latter. It is known that application of a tensile field on a film sandwiched between two parallel plates leads to formation of defects while debonding. These could be fingering due to Saffman-Taylor instability which comes into play when a less viscous fluid displaces a more viscous fluid.[60] Cavitation could also be present because of a large pressure drop in the radial direction.[61, 62] These defects/instabilities, that are absent in squeeze flow,[63] make the elongational flow more complicated. This study can be divided into two parts. In the first part we study the strain response in the directions of the applied orthogonal stress fields. We analyze effect of stress in one direction to strain induced in the other direction. We also investigate effect of orthogonal stress fields on failure in elongational flow. In the second part, we determine the yielding phase diagram under a combination of tensile flow and rotational shear flow to examine validity of Von Mises criterion.

**II.     Materials and Experimental Procedure:**

In this study, we used four different types of soft glassy materials with different microstructures: hair gel, shaving foam, (concentrated) emulsion paint and clay suspension. While the first three were used as purchased, the fourth one was prepared in the laboratory. The first system used in the present study is a commercial hair gel called Brylcreem® Ultra Hard gel. It is a transparent, water based yield stress material. Hair gels are generally composed of a polyelectrolyte polymer dispersed in water. The charge on the polymer prevents it from coiling up and hence it stretches out on application of deformation field. Rheological studies on hair gel samples from the same container showed good reproducibility. Although the qualitative rheological behavior of samples from different containers was identical, the samples demonstrated container to container (batch to batch) variation of various rheological properties. A single container Brylcreem Ultra Hard gel contains 100 ml gel. Since this amount is not sufficient to complete one set of experiments, we have used samples from different containers and have labeled the same as Hair gel -1, 2, 3 and 4.

The second system used is Gillette® regular shaving foam, which is aqueous based foam composed of gas bubbles closely packed in a surfactant solution.[64] Upon application of weak stresses these materials behave as visco-elastic solids. When the yield stress is exceeded visco-plastic behavior is observed. During flow, the bubbles are continuously deformed which leads to mesoscopic structural rearrangements in the material.[65]

The third system used is a paint purchased from Berger Paints® (British Paints). Emulsion paints are generally composed acrylic polymer droplets as a dispersed phase and aqueous continuous phase. These also fall under the category of soft glassy materials which undergo a visco-plastic flow above a yield stress.[66]

The fourth system used is a 3.5 wt. % aqueous Laponite® suspension. Laponite RD used in the present study was purchased from Southern Clay Products, Inc. The Laponite suspension was prepared by the following method: subsequent to drying in the oven for 4 hrs at 120°C, the Laponite RD powder was added slowly to ultra pure water at pH 10 and mixed vigorously using Ultra-Turrax drive T25 until a clear dispersion was obtained. The Laponite suspension is known to undergo ergoditicity breaking with passage of time and converts from a clear liquid dispersion into a soft glassy solid/paste which is able to sustain its own weight over laboratory time scales.[1, 67] A freshly prepared suspension was left undisturbed in a sealed polypropylene bottle for around 3 months. This ensures that the microscopic dynamics of aging has slowed down enough so that no significant aging would occur during the course of the experiments.

Before carrying out each experiment, the sample was shear melted by applying a shear stress greater than the yield stress of the material. Shear melting is necessary to remove the deformation history. We have used the parallel plate geometry (50 mm diameter) of Anton Paar Physica MCR 501 rheometer for all the experiments. It should be noted that in parallel plate geometry radial dependence of stress cannot be neglected. However in order to apply combined stress field as mentioned before parallel plate is the best geometry that can be employed. The free surface of the sample was coated with silicone oil to prevent evaporation and/or contamination with $CO_2$. All the experiments were carried out at 25°C.

## III. Results and Discussion:

In this study we simultaneously apply a constant tensile force and a constant rotational shear stress on the thin films (around 100 µm) of the samples sandwiched between the parallel plates as shown in Fig. 1. Under the application of normal force ($F$) the plates start to separate leading to an increase in tensile strain $\varepsilon$ with time. Tensile strain is defined as $\varepsilon = (d(t)/d_i) - 1$, where $d(t)$ is gap between the plates at time $t$ and $d_i$ is the initial gap. Fig. 2 shows a typical variation of $\varepsilon$ with time under an application of constant normal force $F$ and constant torque $T$ (constant rotational shear stress $\tau_{\theta z}$) for hair gel – 1. The rheometer maintains the normal force at a constant value by a feedback mechanism, which involves controlling the vertical movement of the top plate. Subsequent to the initiation of failure in the film, a sharp decrease in normal force is observed as the rheometer is no longer able to maintain the normal force at a constant value. Hence, we consider the data only up to the point of initiation of failure in all the experiments.

Under application of normal force ($F$), the top plate moves upwards creating a pressure gradient in the radial direction which induces a radial shear stress $\tau_{rz}$. Since gap between the plates is much less than the radius of the fluid film $R$ ($d \ll R$), radial velocity, $v_r$ should also be significantly greater that vertical velocity $v_z$ ($v_r \gg v_z$). In the limit of very small Reynolds number ($\text{Re} \ll 1$), application of lubrication approximation leads to:

$$\tau_{rz} = \frac{3Fd}{2\pi R^3}, \tag{2}$$

while rotational shear stress is given by:

$$\tau_{\theta z} = \frac{3T}{2\pi R^3}, \tag{3}$$

where $T$ is torque. In addition, radial strain rate $\dot{\gamma}_{rz}$ is given by:[46]

$$\dot{\gamma}_{rz} = \frac{VR}{[d(t)]^2}, \tag{4}$$

where $V$ is velocity of top plate ($= d[d(t)]/dt$). Equation 4 can be easily integrated to obtain radial strain:

$$\gamma_{rz} = \frac{2R_0 d_i^{1/2}}{3}\left[\frac{1}{d_i^{3/2}} - \frac{1}{[d(t)]^{3/2}}\right], \tag{5}$$

where $R_0$ is radius of top plate. Furthermore, true rotational strain is given by:

$$\gamma_{\theta z} = \int_0^t \frac{\Omega(t)R(t)}{d(t)} dt, \tag{6}$$

where $\Omega$ is angular velocity of top plate and $\Omega R/d$ is true rotational strain rate ($\dot{\gamma}_{\theta z}$).[42]

As mentioned in the introduction, in the first part of the paper we discuss effect of combined stress fields on the deformation behavior of the material leading to failure. In this part lubrication approximation is applicable only in the limit of small gaps between the parallel plates. However in the neighborhood of failure contribution of elongational flow field dominates. In the second part of the paper phenomenon of yielding is studied which occurs in the limit of small gaps, where lubrication approximation holds. Under such conditions (lubrication approximation) non zero components of the stress tensor are only $\tau_{\theta z}$ and $\tau_{rz}$. Since both these are shear stresses, their superposition is also a shear stress. However, since flow is still normal stress controlled effect of initiation of fingering instability cannot be neglected as discussed later in the paper.

In Fig. 3 we have plotted evolution of tensile strain as a function of time at a constant tensile force ($F = 10$ N) but different rotational shear stresses $\tau_{\theta z}$ on a hair gel – 1 sample. As the sample volume in this test remains constant, radius of the fluid film decreases due to an increase in gap between the plates. This leads to an increase in true values of components of stress tensor even though force $F$ and torque $T$ are constant. Remarkably, we find that imposition of rotational shear stress shifts the evolution of elongational strain to lower times (which means a higher elongational strain for a higher rotational stress at a particular time), even though both the directions are orthogonal to each other. Temporal evolution of elongational strain can be seen to have close to linear slope for small gaps, which increases sharply in the limit of failure. In a previous study, we observed that the elongational strain curves shift to lower times when a higher normal force was applied.[23] Interestingly, in the present case, a higher torque thus not only leads to a higher (true) rotational strain as expected (as shown in the top inset of Fig. 3) but also induces a higher elongational strain.

In Fig. 4 we plot radial strain associated with the data plotted in Fig. 3 but in the limit of small gaps ($d(t)/R_0 \leq 0.05$) by applying lubrication approximation (equation 5). As expected evolution of $\gamma_{rz}$ also shows the same trend as elongational strain. We also investigate the behavior when torque was held constant and the normal force was varied as shown in Fig. 5. Similar to the observation of Fig. 3, we find that the rotational strain is higher for a higher normal force while torque is kept constant. However, increase in the rotational strain with increase in the normal force is small because of a much greater magnitude of $\tau_{\theta z}$ as compared to $\tau_{rz}$ induced due to normal force. We repeat this experiment for other rotational stresses (not shown here) and we observe a small but consistent enhancement in the rotational strain value when the normal force is increased. Azimuthal stress $\tau_{\theta z}$ applied here is either comparable or greater than the yield stress of the material and hence un-jams the system in all directions. Therefore, for the same value of $\tau_{rz}$ (due to normal force), higher elongational strain is observed for greater $\tau_{\theta z}$ as shown in Fig. 3, while for the same value of $\tau_{\theta z}$, greater rotational strain is observed for higher $\tau_{rz}$. In addition, we find that all the rotational strain – time curves and the elongational strain – time curves have a similar curvature. This suggests that, though at different time scales, the same path is followed during the process irrespective of the magnitude of the resultant stress field. We also carried out similar experiments on shaving foam and aqueous Laponite suspension at different normal stresses. Both these materials demonstrate qualitatively similar behavior described in Fig. 3 and 5 suggesting universality of these observations.

Although the constitutive model for yield stress fluid (equation 1) cannot predict failure, it can certainly describe strain induced in the material if we simplify the flow field as represented after applying lubrication approximation (equations 2 and 3). For $\sqrt{\underline{\underline{\tau}}:\underline{\underline{\tau}}/2} \geq \tau_y$, and for nonzero $\tau_{rz}$ and $\tau_{\theta z}$, equation 1 can be analytically solved for a specific case of Herschel-Bulkley model to give:

$$\gamma_{\theta z} = \frac{\left(\tilde{\tau}_{\theta z}\sqrt{1+\left(\tilde{\tau}_{rz}^2/\tilde{\tau}_{\theta z}^2\right)}-1\right)^{1/n}}{\sqrt{1+\left(\tilde{\tau}_{rz}^2/\tilde{\tau}_{\theta z}^2\right)}}\left\{t\left(\frac{\tau_y}{m}\right)^{1/n}\right\} \text{ and} \qquad (7)$$

$$\gamma_{rz} = \frac{\left(\tilde{\tau}_{rz}\sqrt{1+\left(\tilde{\tau}_{\theta z}^2/\tilde{\tau}_{rz}^2\right)}-1\right)^{1/n}}{\sqrt{1+\left(\tilde{\tau}_{\theta z}^2/\tilde{\tau}_{rz}^2\right)}}\left\{t\left(\frac{\tau_y}{m}\right)^{1/n}\right\}, \qquad (8)$$

where $\tilde{\tau}_{\theta z} = \tau_{\theta z}/\tau_y$ and $\tilde{\tau}_{rz} = \tau_{rz}/\tau_y$. We plot evolution of $\gamma_{rz}$ given by equation 8 as a function of normalized time for a constant value of $\tilde{\tau}_{rz}$ but different values of $\tilde{\tau}_{\theta z}$ in the inset of Fig. 4. It can be seen that Herschel Bulkley model qualitatively describes the experimental behavior wherein increase in $\tau_{\theta z}$ is observed to shift evolution of $\gamma_{rz}$ to smaller times even though $\tau_{rz}$ is constant. Similarly behavior described in Fig. 5 can be explained by equation 7. It should be noted that in the limit of $\tilde{\tau}_{\theta z} \gg \tilde{\tau}_{rz} > 1$ equation 8 reduces to: $\gamma_{rz} = \tilde{\tau}_{rz}\tilde{\tau}_{\theta z}^{(1-n)/n}\{t(\tau_y/m)^{1/n}\}$. For $n=1$, that is for Bingham model, $\gamma_{rz}$ becomes independent of $\tilde{\tau}_{\theta z}$ in this limit. Experimental data shown in Fig. 4, however, shows that evolution of $\gamma_{rz}$ shifts to smaller times without showing any sign of shift getting truncated. Therefore the experimental behavior described in Fig. 4 is better predicted for $n<1$, that is Herschel Bulkley model than Bingham model. Interestingly MCT based schematic model proposed by Brader and coworkers[45, 59] demonstrates various features of Herschel Bulkley model, which we believe should also enable the former to qualitatively predict the experimental behavior shown in Fig. 3 to 5.

  Since the soft jammed material has a three dimensional structure, it is expected that the invariant of the stress tensor governs the level of unjamming and the distance of its value from the yield stress decides the state of the material.[46] Fig. 6 shows the time to failure $t_f$, which is the instant of time at which failure begins in the film marked by a sharp decrease in the normal force (Fig. 2), as a function of second invariant of (true) stress tensor under application of different normal forces. Interestingly, we observe that $t_f$ does not depend only on normal force but also on magnitude of $\tau_{\theta z}$. Furthermore, the time to failure can be seen to decrease with increase in the normal force for approximately the same invariant of stress tensor. In addition $t_f$ decays with close to reciprocal dependence on the invariant of stress tensor. We believe that this behavior is also due to a greater unjamming of the material caused by a progressively greater $\tau_{\theta z}$ leading to failure at lower times, even though the applied normal force is identical. It has been reported that the time to failure follows an inverse dependence on the applied normal stress.[68] Interestingly, in the present case, where more than one stress fields is applied, time to failure decays inversely with the invariant of the stress tensor.

As mentioned in the introduction section, owing to contraction of cross sectional area when two plates are pulled apart, low viscosity fluid (air) pushes high viscosity fluid (sample) thereby causing fingering due to Saffman Taylor instability. We also observe this phenomenon in our experiments. In Fig. 7 we show photo of the pattern formed on the top plate of rheometer for shaving foam immediately after the separation of two plates (failure). It can be seen that in the limit of flow field dominated by normal force acting on the top plate (Fig. 7 a) sample indeed undergoes fingering instability. However with increase in rotational shear stress fingers tend to distort in the azimuthal direction, and in the limit of dominant rotational flow fingers disappear completely. We also observe similar behavior for all the soft materials explored in this work. Observation of fingering instability also suggests that radial and elongational strains are not uniform in the azimuthal direction in the limit of tensile stress dominated flow. Similarly calculation of true stress will also get affected in that limit. Therefore it is necessary to keep these issues in mind while analyzing the data when flow is dominated by normal force on the top plate.

Soft glassy materials undergo a transition from solid to liquid state when the second invariant of applied stress field exceeds the yield stress. As discussed before, application of stress facilitates movement of trapped particles out of their cages causing a structural breakdown which sets flow in the material. The yielding curve in the case of both the stresses ($\tau_{\theta z}$ and $\tau_{rz}$) applied simultaneously would constitute various combinations which could lead to yielding in the material. In order to determine the yielding curve, we maintain the normal tensile force at a constant value and apply a shear stress ramp. The value of rotational shear stress at which yielding takes place (the point at which the flow begins accompanied by a sharp decrease in viscosity) corresponds to the critical $\tau_{\theta z,y}$ for that force or radial stress. However, since various materials studied in this work are also thixotropic, where the yield stress is a function of deformation history, the yield stress may depend upon the rate at which the shear stress is varied. Therefore the yield stress values obtained at different shear stress ramps could be different. To investigate this point, we carried out a set of experiments on a hair gel sample 'hair gel-2', in which we vary the rate of change of shear stress at the same normal force and determine the yield stress. It can be seen from Fig. 8 that the yield stress values obtained for the same normal force are weakly dependent (around 20 to 30 % variation around mean) on the rate of variation of the shear stress in the explored window.

Fig. 9 shows viscosity as a function of $\tau_{\theta z}$ for hair gel – 3 for various normal forces. It can be clearly seen that as the normal force value increases the yield stress shifts towards lower values. We performed this procedure for four more soft glassy materials namely hair gel – 4, shaving foam, emulsion paint and Laponite suspension. Fig. 10 shows jamming phase diagram under various combinations of true values of radial and rotational shear stresses at which yielding takes place, normalized with the yield stress obtained by purely rotational experiments. The material is in the solid state in the region inside the phase boundary (curve formed by these points), while in the liquid state outside it. Closer inspection of the data suggests that there is significant scatter in the limit of normal force dominated experiments. In addition Laponite suspension shows slightly smaller value of yield stress in normal force dominated flow compared to purely rotational flow. We believe that this might be originating from initiation of defects such as fingering shown in Fig. 7. In addition even calculation of true radial stress in this limit is expected to be affected by the same. However, except this deviation most of the points do lie on the line representing Von Mises criterion thereby validating the same. Ovarlez et al.,[46] validated the Von Mises criterion under combined squeeze and rotational flow fields (by carrying out rate controlled experiments), on the other hand, we observe the Von Mises criterion is followed closely when tensile and rotational stresses act simultaneously. This result is interesting because, an elongational flow in the present context is not merely a reversal of direction of squeeze but is much more complex compared to a squeeze flow because of a possibility of defect formation like fingering/crack formation, cavitation due to a large pressure gradient in the radial direction, etc., which can in principle alter the behavior.

**IV. Conclusions:**

We study deformation and yield behavior of various soft glassy materials with very different microstructures such as commercial hair gel, emulsion paint, shaving foam and clay suspension, when acted upon simultaneously by constant tensile stress and shear stress. The paper is divided into two parts. In the first part we analyze deformation behavior of various soft jammed materials when acted upon by two orthogonal stress fields. Interestingly, we observe that stress applied in one direction affects the strain induced in the other directions as well. Application of tensile flow field eventually leads to failure in the sample. In addition, the time to failure is observed not to depend only on the normal force acting on the sample but also on the rotational shear stress applied to the sample. Typically for a given normal force, application of greater magnitude of rotational shear stress is observed to induce failure over a shorter duration. In addition, we observe that the strain response, under various combinations of tensile and rotational shear stresses, is self-similar in nature suggesting a shift only in the time - scales and not the way material gets deformed. We believe that the observed deformation and failure behavior is due to overall unjamming of the system caused by stress applied in one direction whose effect gets compounded when stress is applied in other direction. We solve Herschel-Bulkley model when acted upon by two stress fields analytically. Remarkably Herschel-Bulkley model qualitatively predicts experimentally observed deformation behavior of soft jammed materials well. In the second part of the paper we plot jamming phase diagram for solid (jammed state) – liquid (flowing state) yielding transition by simultaneously varying magnitudes of normal force and rotational shear stress. Typically, the yielding behavior is studied by applying a rotational shear stress ramp for every applied normal force. Over the explored shear stress ramp rates, yield stress is observed to be practically independent of ramp rates. We observe scatter in the yield stress values in the limit of flow field dominated by normal force on the top plate, which might be due to fingering instability. However, except this deviation, yield stress is indeed observed to be an invariant of various stress fields acting on the material validating the Von Mises criterion.

**Acknowledgement:** Financial support from Department of Science and Technology through IRHPA scheme is greatly acknowledged.

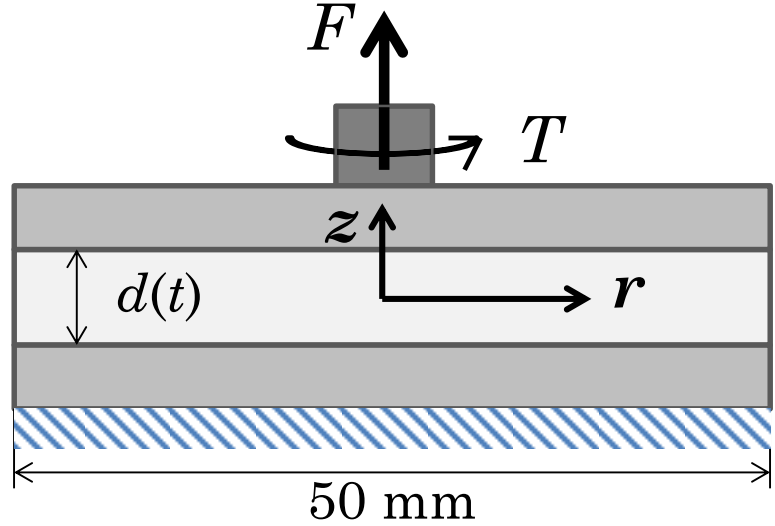

**Fig. 1** Schematic parallel plate geometry employed to carry out the experiments. We use the polar coordinate system. Azimuthal ($\theta$) direction (not shown) is orthogonal to both $z$ and $r$ directions. Rotational shear stress ($\tau_{\theta z}$) is due to force in $\theta$ direction acting on $r$ - $\theta$ surface while radial shear stress ($\tau_{rz}$) is due to force in $r$ direction acting on $r$ - $\theta$ surface.

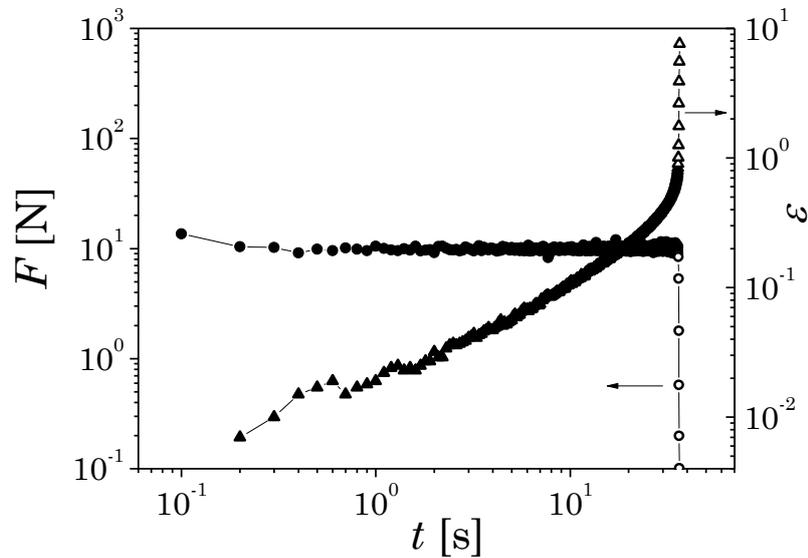

**Fig. 2** Applied normal force $F$ (circles) and elongational strain $\varepsilon$ (triangles) as a function of time $t$ for a hair gel – 1 sample. In this experiment a constant rotational shear stress, $\tau_{\theta z}$ = 100 Pa is also applied simultaneously along with the normal force. The filled symbols represent the data points corresponding to a constant value of $F$ = 10 N. A rapid drop is observed in $F$ subsequent to the initiation of failure in the film shown with open symbols.

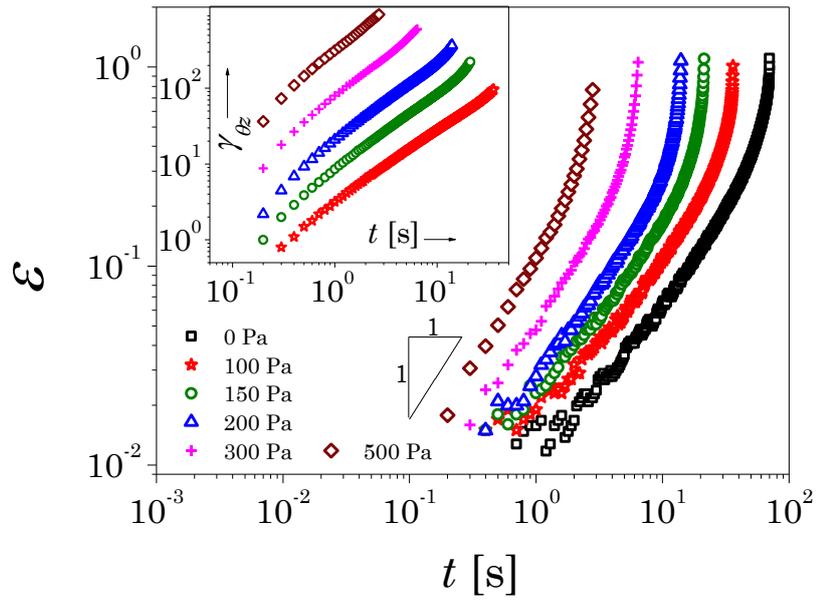

**Fig. 3** Elongational strain as a function of time for a constant normal force $F = 10$ N and various rotational shear stresses for hair gel - 1. The inset shows the corresponding true rotational strain $\gamma_{\theta z}$ for each rotational shear stress $\tau_{\theta z}$ as a function of time.

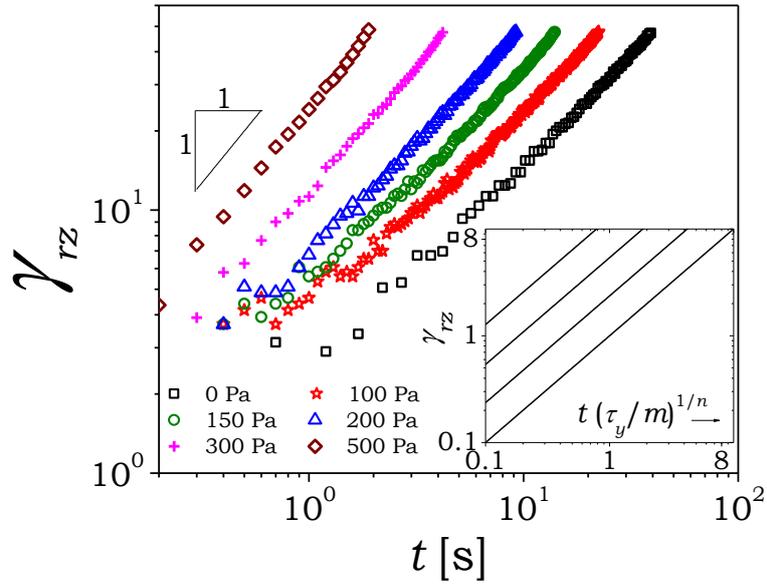

**Fig. 4** Radial shear strain $\gamma_{rz}$ for the same data shown in Fig. 2 plotted as a function of time for constant normal force of 10 N ($\tau_{rz}$ = 30.6 Pa) but different torques (initial $\tau_{\theta z}$ mentioned in legend). The inset shows prediction of Herschel Bulkley model (equation 8) wherein $\gamma_{rz}$ is plotted as a function of normalized time for $n$ =0.5, $\tau_{rz}/\tau_y$ =2, and varying $\tau_{\theta z}/\tau_y$ =0, 2, 4, 8 (from right to left).

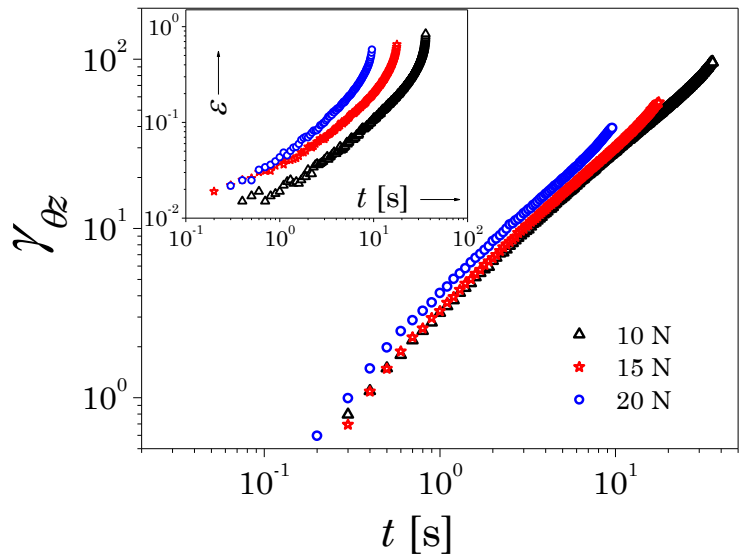

**Fig. 5** True rotational shear strain is plotted as a function of time for a constant torque (initial shear stress $\tau_{\theta z}$ = 100 Pa) and various normal forces for hair gel - 1. The inset shows the corresponding elongational strain $\varepsilon$ for each normal force $F$ as a function of time.

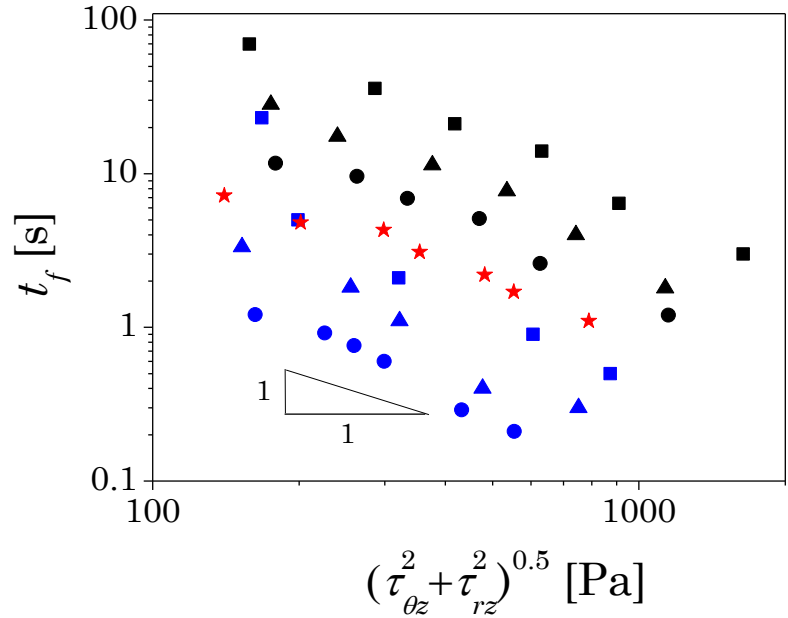

**Fig. 6** Time to failure as a function of the invariant of true stress tensor corresponding to various normal forces: 5 N ($\tau_{rz}$ = 15.3 Pa, stars), 10 N ($\tau_{rz}$ = 30.6 Pa, squares), 15 N ($\tau_{rz}$ = 45.9 Pa, triangles) and 20 N ($\tau_{rz}$ =61.2 Pa, circles) and various shear stresses ($\tau_{\theta z}$ =0 to 500 Pa). Black symbols represent hair gel - 1, red symbols represent shaving foam, and blue symbols represent Laponite suspension.

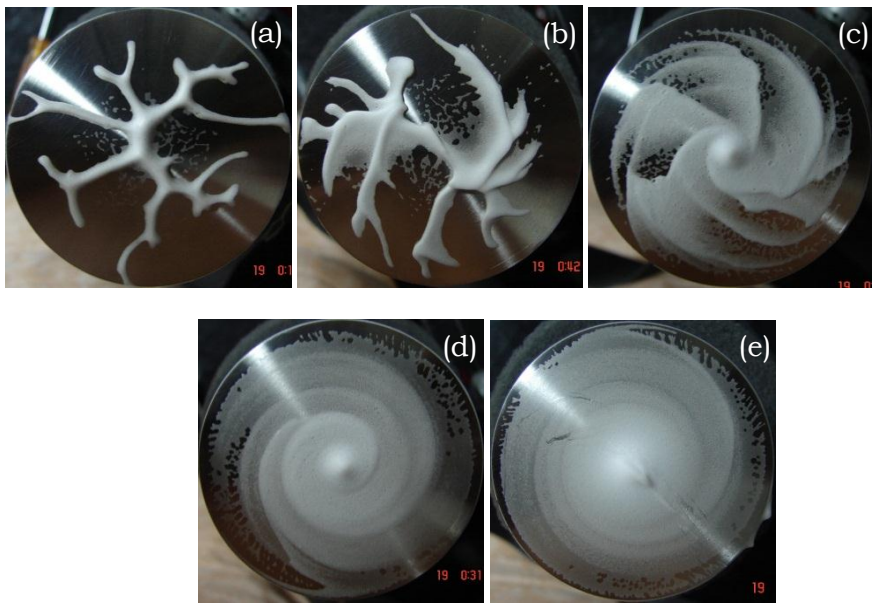

**Fig. 7** Patterns formed on the upper plate after completion of separation for foam: (a) $F$ =5 N, $\tau_{\theta z}$ = 0 Pa; (b) $F$ =5 N, $\tau_{\theta z}$ = 50 Pa; (c) $F$ =5 N, $\tau_{\theta z}$ = 100 Pa; (d) $F$ =5 N, $\tau_{\theta z}$ = 200 Pa and (e) $F$ =5 N, $\tau_{\theta z}$ = 300 Pa. The initial gap between the plates $d_i$ =100 µm and the radius of the top plate $R$ =25 mm in all the experiments. As the failure

is cohesive, the patterns formed on the lower plate are a mirror image of that on the upper plate.

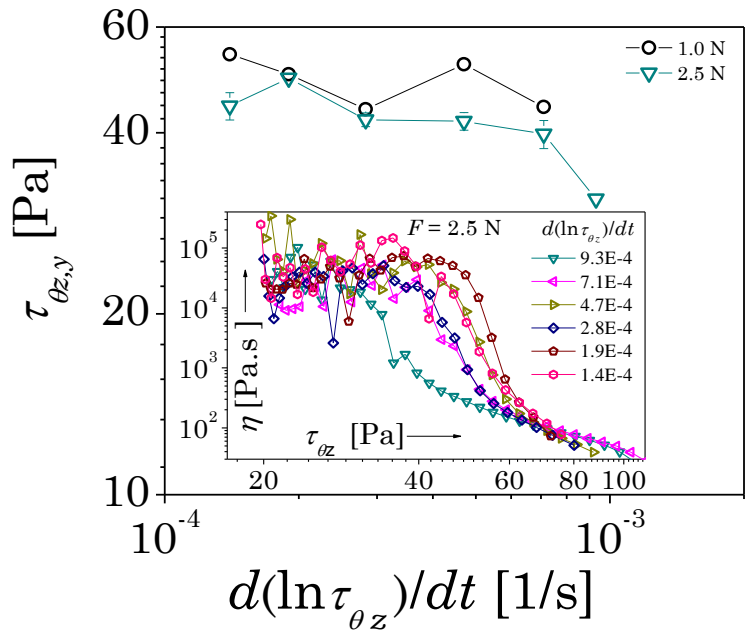

**Fig. 8** Yield stress $\tau_{\theta z,y}$ is plotted against rate of variation of stress $d(\ln \tau_{\theta z})/dt$ for Hair gel - 2. The inset shows the viscosity as a function of the rotational shear stress for various rates of variation of the rotational shear stress but at a constant normal force of 2.5 N.

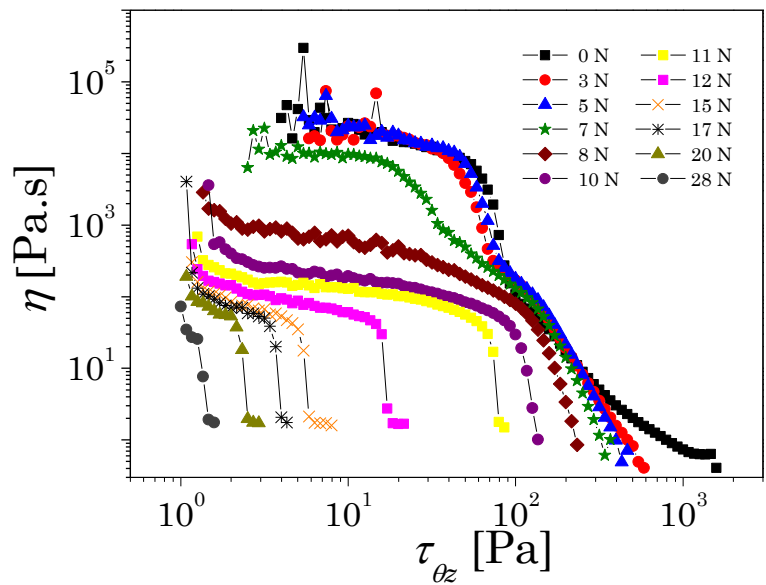

**Fig. 9** Viscosity $\eta$ as a function of rotational shear stress under the application of various tensile normal forces for hair gel - 3.

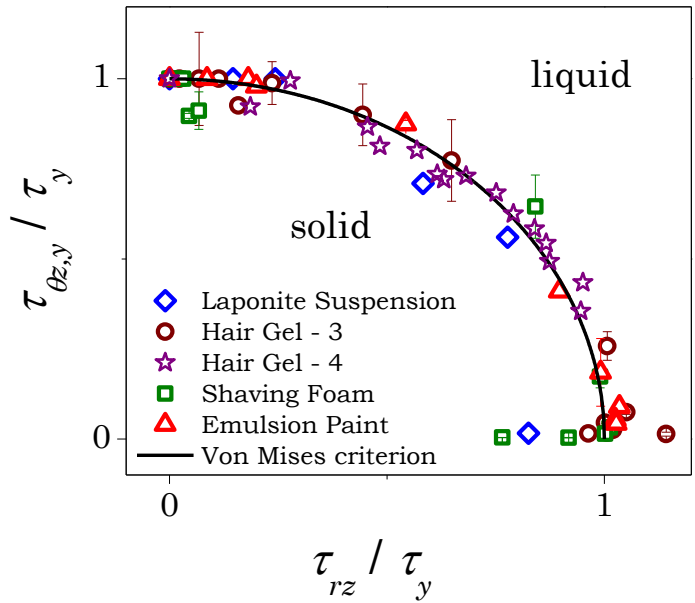

**Fig. 10** The yielding phase diagram for various fluids (Laponite suspension: $\tau_y$=80 Pa, hair gel - 3: $\tau_y$=135 Pa, hair gel – 4: $\tau_y$=18.3 Pa, shaving foam: $\tau_y$=68 Pa, emulsion paint: $\tau_y$=35 Pa) where $\tau_y$ is the yield stress in simple shear. The symbols represent various combinations of normalized rotational shear stress and normalized radial shear stress at which the yielding takes place. The behaviour can be explained well with the Von Mises criterion which is shown as a solid line.